
\magnification=1200 \vsize=25truecm \hsize=16truecm \baselineskip=0.6truecm
\parindent=1truecm \nopagenumbers \font\scap=cmcsc10 \hfuzz=0.8truecm
\def\uup{\overline u}
\def\udo{\underline u}
\def\Aup{\overline A}
\def\zup{\overline z}

\def\vup{\overline v}
\def\vdo{\underline v}
\def\fup{\overline f}
\def\fdo{\underline f}
\def\tup{\overline \tau}
\def\tdo{\underline \tau}
\def\fuup{\overline {\fup}}
\def\guup{\overline {\gup}}
\def\gup{\overline g}
\def\gdo{\underline g}

\def\dhdot{\!\cdot\!}

\null \bigskip  \centerline{\bf BILINEAR DISCRETE PAINLEV\'E-II AND  ITS
PARTICULAR SOLUTIONS}
\vskip 2truecm
\centerline{{\scap J. Satsuma}}
\centerline{\sl Department of Mathematical Sciences}
\centerline{\sl University of Tokyo}
\centerline{\sl 3-8-1 Komaba, Meguro-ku, Tokyo 153, Japan}
\bigskip
\centerline{{\scap K. Kajiwara}}
\centerline{\sl Department of Electrical Engineering}
\centerline{\sl Doshisha University}
\centerline{\sl Tanabe, Kyoto 610-03, Japan}
\bigskip
\centerline{\scap B. Grammaticos}
\centerline{\sl LPN, Universit\'e Paris VII}
\centerline{\sl Tour 24-14, 5${}^{\grave eme}$\'etage}
\centerline{\sl 75251 Paris, France}
\bigskip
\centerline{{\scap J. Hietarinta}}
\centerline{\sl Department of Physics}
\centerline{\sl University of Turku}
\centerline{\sl 20500 Turku, Finland}
\bigskip
\centerline{\scap A. Ramani}
\centerline{\sl CPT, Ecole Polytechnique}
\centerline{\sl CNRS, UPR 14}
\centerline{\sl 91128 Palaiseau, France}

\vskip 3truecm \noindent Abstract \smallskip
\noindent By analogy to the continuous Painlev\'e II equation, we present
particular solutions
of the discrete Painlev\'e II (d-P$\rm_{II}$) equation. These solutions are
of rational and
special function (Airy) type. Our analysis is based on the bilinear
formalism that allows us to
obtain the $\tau$ function for d-P$\rm_{II}$. Two different forms of
bilinear d-P$\rm_{II}$ are
obtained and we show that they can be related by a simple gauge
transformation.

\vfill\eject

\footline={\hfill\folio} \pageno=2

\noindent The (continuous) Painlev\'e equations are known as paradigms of
integrability [1].
Still, their integrability is of a special type, in the sense that it does
not lead to the solution
being written in terms of known functions. In fact, the general solution
can be obtained only
through IST methods (which, in essence, reduce the Painlev\'e equations to
a linear
integrodifferential equation) [2]. Although the general solution to the
Painlev\'e equations is
essentially transcendental, simple solutions do also exist. In particular
for those
Painlev\'e equations that contain free parameters it is possible to find
values of these
parameters for which a solution can be obtained in terms of special
functions [3]. It must be
pointed out, though, that this `elementary' solution does not possess the
full dimensionality of
integration constants. Typically, the Painlev\'e equations have (for
special values of their
parameters) solutions involving one integration constant and which are
given in terms of special
functions or solutions that do not involve any integration constant and are
usually rational
expressions.

Let us illustrate this in the case of the second Painlev\'e equation
P$\rm_{II}$:
$$w''=2w^3+xw+\alpha\eqno(1)$$ It is well known that whenever the parameter
$\alpha$ is half
integer P$\rm_{II}$ possesses solutions that can be expressed in terms of
Airy functions while
for integer $\alpha$ P$\rm_{II}$ has solutions of rational type. Thus when
$\alpha =1/2$ we have
$w=-A'/A$ with
$A$ solution of $A''+{x\over 2}A=0$. The simplest rational solution is
obtained for
$\alpha=0$ in which case we find $w=0$. The construction of solutions for
higher values of the
parameter can be based on the auto-B\"acklund transform of P$\rm_{II}$ [4],
that relates the
solutions of P$\rm_{II}$ corresponding to values $\alpha$ and $\alpha+1$:
$$w(\alpha+1)=-w(\alpha)-{1+2\alpha\over
2w^2(\alpha)+2w'(\alpha)+x}\eqno(2)$$ The parity relation
$w(-\alpha)=-w(\alpha)$ allows us to reach the negative values of the
parameter $\alpha$. Similar
results exist for the higher Painlev\'e equations.

 An important result that has been obtained
recently concerning these rational solutions. It was shown in [5] that they
can be expressed in
terms of $\tau$ functions in the form of Wronskian determinants:
$$\tau_N=\left| \matrix{
a_0 & a_1 &\cdots & a_{N-1}\cr
a_1 & a_2 &\cdots & a_{N}\cr
\vdots & \vdots & \ddots &\vdots\cr
a_{N-1} & a_{N} &\cdots & a_{2N-2}\cr}\right|\eqno(3)$$
where the $a_k$'s satisfy
$$a_k=a'_{k-1}+\sum_{j=0}^{k-2}a_ja_{k-j-2}\ ,\eqno(4)$$
with
$$a_0=x\ ,a_1=1\eqno(5)$$
Equations(3) and (4) uniquely determine $a_k$. We must point out here that
the $a_k$'s
satisfy also a {\sl linear} recursion relation,
$$a'_k=2(k-1)a_{k-2}\eqno(6)$$
which however does not suffice for their complete determination.
 The nonlinear variable $w$ is related to the $\tau$-function though:
$$w=\left(\log {\tau_{N+1}\over \tau_N}\right)'\eqno(7)$$
and satisfies P$\rm_{II}$ in the form:
$w''=2w^3-4xw+4(N+1)$, which is equivalent to (1) up to a scaling of both
$w$ and $x$.

In this letter we shall focus on the discrete equivalent of these results.
As is well known the
Painlev\'e equations possess discrete analogs [6,7] that are integrable in
a sense reminiscent of
the continuous case: isomonodromy based on the existence of a Lax pair [8].
Special solutions do
also exist. In a series of publications [9,10,11] we have analysed this
problem and obtained
results for the d-P's from II to V. In what follows we shall return to
d-P$\rm_{II}$:
$$\uup+\udo={zu-a\over 1-u^2}\eqno(8)$$ where we use the notation $u\equiv
u_n$, $\uup\equiv
u_{n+1}$ and $\udo\equiv u_{n-1}$. Note that $u$ has the parity
$u(a)=-u(-a)$ in close analogy
to the continuous case. What is known about this equation? In [8] we have
obtained the (discrete)
Miura and auto-B\"acklund transformations. The former writes:
$$v=(1-\uup)(u+1)-z/2-\delta /4\eqno(9)$$ and transforms d-P$\rm_{II}$ to
d-P$\rm_{34}$:
$$(\vup+v)(v+\vdo)={4v^2-m^2\over v+z/2+\delta /4}\eqno(10)$$ where
$m=a-\delta/2$ and
$\delta\equiv\zup -z$. The discrete Miura transform (9) is complemented by:
$$u={v-\vdo-m\over v+\vdo}\eqno(11)$$ This means that eliminating $v$ or
$u$ between (9) and (11)
we can find (8) or (10) respectively. The auto-B\"acklund transformation
relating $u(a)$ to
$u(a-\delta)$ writes:
$$u(a-\delta)=-u(a)-{(2a-\delta)(1+u(a))\over 2(u(a)+1)(1-\uup
(a))-z-a}\eqno(12)$$ Airy-type
solutions were shown to exist. For instance at $a=\delta/2$ we have
$u=-1+\Aup /A$ where $A$ is a
solution of the discrete Airy equation:
$\overline{\Aup}-2\Aup+(z/2+\delta/4)A=0$. The basic rational solution, as
in the continuous
case, is $u=0$ for
$a=0$. From these `seed' solutions one can construct higher ones using the
B\"acklund transform
(12).

A different approach was presented in [12] where we obtained the general
Airy-type solution
of d-P$\rm_{II}$ in terms of Casorati determinants. The essential tool in
that approach is the
bilinear formalism: the solution to the d-P$\rm_{II}$ equation is expressed
as a ratio of
$\tau$ functions each of which is a Casorati determinant. However, this
approach is a `top-down'
one, starting from the determinantal form of the solution and using
determinantal (Pl\"ucker)
identities in order to obtain the equation. This may not be always
convenient. In this letter we
shall present a different treatment which is also based entirely on the
bilinear formalism but
starts with the equation and proceeds to the construction of the solution.
This approach is
particularly convenient for the rational solutions of d-P$\rm_{II}$ as we
will show below.

Let us start with the bilinear form of d-P$\rm_{II}$. The bilinear forms of
the continuous
Painlev\'e equations have been presented in [13]. It was shown there that
starting from the
equation (1) and introducing $F$, $G$ through $w=(\ln F/G)_x$ one can
obtain a bilinearization
in the form:
$$D^2 F\dhdot G=0\eqno(13a)$$
$$(D^3-xD-\alpha)F\dhdot G=0\eqno(13b)$$ where $D$ is the Hirota operator
acting on a
dot-product. Its precise definition is the following:
$DF\cdot
G=(\partial_x-\partial_{x'})F(x)G(x')|_{x=x'}=F'(x)G(x)-F(x)G'(x)$. The
quantities
$F$ an $G$ are indeed $\tau$ functions for the continuous P$\rm_{II}$. In
order to construct the
discrete analog to (13) one must recall the relation between the parameters
of the continuous
P$\rm_{II}$ (1) and those of the discrete one (8). In fact in order to
obtain the continuous
limit of d-P$\rm_{II}$ we must take
$a=\epsilon^3\alpha$,
$z=2+\epsilon^2x$ and let $\epsilon\to 0$ while $u$ becomes $w$. We shall
not present here a
fully systematic approach of the bilinearization of d-P$\rm_{II}$. It is
clear that the
continuous $D$'s must be replaced by the discrete shift operators $e^D$ the
action of which is
given by $e^Df\dhdot g=f(n+1)g(n-1)$. To make a long story short, the
bilinear form of
d-P$\rm_{II}$ reads just:
$$(\cosh D-1)f\cdot g=0\eqno(14a)$$
$$(\sinh 2D-z\sinh D-a)f\cdot g=0\eqno(14b)$$ Equations (14) can be written
in a more explicit way
as:
$$\fup\gdo+\fdo\gup-2fg=0\eqno(15a)$$
$$\fuup\underline{\gdo}-\guup\underline{\fdo}-z(\fup\gdo-\fdo\gup)
-2afg=0\eqno(15b)$$ The
relation of the nonlinear variable $u$ to the $f$,$g$ is
$$u={\fup\gdo\over fg}-1=1-{\fdo\gup\over fg}\eqno(16)$$  The resemblance
to the continuous case
is striking. In fact, one can check that at the continuous limit (14) goes
over to (13). Moreover,
starting with (9) one can obtain a nonlinear equation for the quantity $u$
that turns out to be
exactly the d-P$\rm_{II}$ (8). In order to assign a particular meaning to
the quantities $f$ and
$g$ it is interesting to examine the Miura transform (9). It writes:
$$v={\guup\gdo\over \gup g}-z/2-\delta/4\eqno(17)$$  Next we use the
symmetry property
$u(a)=-u(-a)$ rewritten in terms of the parameter $m$: we have
$u(m+\delta /2)=-u(-(m+\delta)+\delta /2)$. Thus the Miura transform (9)
can be used to construct not
only the solution $v$ corresponding to $m$ but also the solution $w$
corresponding to $-(m+\delta)$
and thanks to the invariance of (10) with respect to $m\to -m$ this
solution coincides with the
one at  $(m+\delta)$. Implementing the Miura transform we find:
$$y={\fuup\fdo\over \fup f}-z/2-\delta/4\eqno(18)$$ At this stage the
situation becomes clear:
$g$ is related to $m=a-\delta /2$ while $f$ goes with $m'=a+\delta /2$. In
fact the $g$ and $f$
are $\tau$ functions corresponding to two consecutive values of the
parameter $a$ with
difference $\delta$.

We now turn to the construction of the rational solutions of (14). As we
have already explained in
[9] equation (8) possesses rational solutions that do not contain any free
parameter for values
of the parameter $a$ that are integer multiples of $\delta$. Thus when
$a=0$ we find
$u=0$ as a solution, while for $a=\delta$ we obtain $u=\delta/(z-2)$. In
order to construct the
rational solutions for higher values of $a$ one can use the auto-B\"acklund
transform(12). However
the use of $\tau$ functions leads to this result in a more natural way. Let
us first rewrite
d-P$\rm_{II}$ so as to introduce directly the parameters related to the
continuous limit
($\delta=-\epsilon^3$):
$$\uup+\udo={u(2-\epsilon^3n)+\epsilon^3(N+1)\over 1-u^2}\eqno(19)$$ In
terms of the
$\tau$-functions, $u$ can be written simply as:
$$u={{\overline\tau_{N+1}}{\underline\tau_{N}}\over\tau_{N+1}\tau_{N}}-1
\eqno(20)$$ We start
with $\tau_0=1$ (and also $\tau_{-1}=1$) and $\tau_1=n$. We then use the
auto-B\"acklund
transformation in order to compute the $\tau$'s for higher $N$'s. We obtain
thus ($\nu\equiv
\epsilon^{-3/2}$):
$$\tau_2=n(n^2-1)-4\nu^2\eqno(21)$$
$$\tau_3=n^2(n^2-1)(n^2-4)-4n(5n^2-8)\nu^2-80\nu^4$$ and so on. What is
really interesting is
that we can, just as in the continuous case, cast these
$\tau$-functions in the form of a determinant. We find thus for $\tau_3$
the expression:

$$\tau_3=\left|\matrix{&n&2\nu&n(n+2)\cr &2\nu&n^2-1&8\nu(n+{1\over
2}-{3\over 4}\sigma)\cr
&n(n-2)&8\nu(n-{1\over 2}+{3\over
4}\sigma)&2n(n^2-4)+20\nu^2\cr}\right|\eqno(22)$$ where
$\sigma$ is just a sign, $\sigma=\pm 1$. Higher $\tau$'s can be (and have
been) computed and
cast in the form of a Casorati determinant.
$$\tau=\left|\matrix{&a_0(n)&a_1&a_2(n+1)&a_3^-(n+2)&\dots\cr
&a_1&a_2(n)&a_3^+(n)&a_4(n+1)&\dots\cr
&a_2(n-1)&a_3^-(n)&a_4(n)&a_5^+(n)&\dots\cr
&a_3^+(n-2)&a_4(n-1)&a_5^-(n)&a_6(n)&\dots\cr
&\vdots&\vdots&\vdots&\vdots&\ddots\cr}\right|\eqno(23)$$
 where the following relation holds $a_k^{\pm}(n,\sigma)=a_k(n\pm{1\over
2},\pm\sigma)$ and:
$$a_0=n$$
$$a_1=2\nu$$
$$a_2=n^2-1$$
$$a_3=8\nu(n-{3\over 4}\sigma)$$
$$a_4=2n(n^2-4)+20\nu^2\eqno(24)$$
$$a_5=32\nu\left((n+{3\over 4}\sigma)^2-{19\over 16}\right)$$
$$a_6=5(n^2-1)(n^2-9)+200\nu^2n$$
$$a_7=128\nu\left((n-{3\over 4}\sigma)^3-{73\over 16}(n-{3\over
4}\sigma)+{15\over
4}\nu^2-{3\over 64}\sigma\right)$$
$$a_8=14n(n^2-4)(n^2-16)+4\nu^2(350n^2-557)$$ What is lacking at the
present stage is the
equivalent of relations (3,4,5) that allows the construction of the matrix
elements. It is not
very difficult to obtain the equivalent of the (linear) differential
relation (5) although some
complications appear because of the parity dependence of the matrix
elements. We have, for even $k$:
$$a_k(n+1)-a_k(n-1)=4(k-1)a_{k-2}(n)\eqno(25a)$$
while for odd $k$ we find:
$$a_k(n+1,\pm\sigma)-a_k(n-1,\pm\sigma)=4(k-1)a_{k-2}(n,\mp\sigma)\eqno(25b)
$$

 However the
equivalent of the nonlinear relation (3) has not been found in the discrete
case. Still, its
existence would lead only to a practical simplification. The important
point is that the $\tau$
functions (which are the fundamental objects) can be computed in an
algorithmic way. In fact,
instead of using the B\"acklund for $u$, equation (12), we can derive a
B\"acklund relation
directly for the
$\tau$-functions:
$$\tau_{N+1}{\tup}_{N-1}=2\nu^2{\overline{\tup}}_N\tdo_N+\tup_N\tau_N(n-N-2
\nu^2)\eqno(26)$$
$$\tup_{N+1}{\tau}_{N-1}=2\nu^2{\overline{\tup}}_N\tdo_N+\tup_N\tau_N(N+n+1-
2\nu^2)\eqno(27)$$
Next we turn to an important question concerning the comparison of the
present results with
those of our work on the Airy-type solutions of d-P$\rm_{II}$. In [12], we
have obtained the
following bilinear expressions:
$$\tau^{n-1}_{N+1}{\tau}^{n+2}_{N-1}=\tau^{n-1}_N\tau^{n+2}_N-\tau^{n}_N\tau
^{n+1}_N
\eqno({\rm K}15)$$
$$\tau^{n+2}_{N+1}{\tau}^{n+1}_{N}-2\tau^{n+1}_{N+1}\tau^{n+2}_N+(pn+q)
\tau^{n}_{N+1}
\tau^{n+3}_N=0\eqno({\rm K}16)$$
$$\tau^{n+1}_{N+1}{\tau}^{n+2}_{N-1}=-(p(n+2N)+q)\tau^{n+2}_N\tau^{n+1}_N+(p
n+q)\tau^{n}_N
\tau^{n+3}_N\eqno({\rm K}17)$$
and with the dependent variable transformation
$$w_n={\tau^{n+1}_{N+1}{\tau}^{n}_{N}\over
\tau^{n}_{N+1}{\tau}^{n+1}_{N}}-1\eqno({\rm K}18)$$
we obtain the standard nonlinear form of d-P$\rm_{II}$:
$$w_{n+1}+w_{n-1}={(2pn+(2N-1)+2q)w_{n}-(2N+1)p\over 1-w_{n}^2}\eqno({\rm
K}19)$$
We remark that equation (K19) corresponds to $a=(2N+1)p$ while the $a$ of
equation (19) is
$-\epsilon^3(N+1)$. The interpretation of this value of $a$ is the
following. Equation (K19), and
also (K15-18), are written specifically for the case of Airy-type
solutions, i.e. half-integer
values of $a/\delta$ and the $\tau_{N}$ of (K18) is associated to
$a=(2N+1)p$. In the present
paper, the formalism of (15ab) is quite general and does not depend on the
type of
solution considered. (Only when rational solutions are considered does $N$
have to be taken
as an integer).  If we wish to compare the bilinear
expressions (K15-17) and (15ab,26,27) we must take into account the shift
in $N$. The simplest way
to do this is to rewrite  (K19) as:
$$w_{n+1}+w_{n-1}={(2p(n+N)+2q)w_{n}-2(N+1)p\over 1-w_{n}^2}\eqno(K19')$$
Comparing (K19$'$) to (19) we obtain $p=-\epsilon^3/2=-1/2\nu^2$ and
$q=1+N/2\nu^2$.
Next, let us remark that there exists a further small difference in the
definitions of the
$\tau$'s in the two papers: what is called $\tau_M(m)$ in [12] corresponds
in this paper to a
$\tau$-function, which we will here denote by $\kappa_M(m+M-N-1)$, i.e.
with the same lower
index but an upper index $(m+M-N-1)$, so that (K18) becomes identical to
our equation
(20). We can now rewrite (K15-17) in terms of the $\kappa$'s:
$$\kappa^{n-1}_{N+1}{\kappa}^{n}_{N-1}=\kappa^{n-2}_N\kappa^{n+1}_N-
\kappa^{n-1}_N
\kappa^{n}_N\eqno(28)$$
$$\kappa^{n+2}_{N+1}{\kappa}^{n}_{N}-2\kappa^{n+1}_{N+1}\kappa^{n+1}_N+\Bigl
(1+{N-n\over 2\nu^2}\Bigr)
\kappa^{n}_{N+1}\kappa^{n+2}_N=0\eqno(29)$$
$$\kappa^{n+1}_{N+1}{\kappa}^{n}_{N-1}=-\Bigl(1-{N+n+1\over
2\nu^2}\Bigr)\kappa^{n+1}_N
\kappa^{n}_N+\Bigl(1+{N-n\over
2\nu^2}\Bigr)\kappa^{n-1}_N\kappa^{n+2}_N\eqno(30)$$
Clearly, equation (29) involving only two consecutive lower indices is
analogous to our equation (15a). The difference in form between these two
equations can
be explained by a gauge transformation between our $\tau$'s and the
$\kappa$'s.
Defining a gauge function $h$ we put
$$\kappa^{m}_{M}=h^{m}_{M}\tau^{m}_{M}\eqno(31)$$
In order for $w_n$ to coincide with $u_n$ we must have
${h^{n+1}_{N+1}h^{n-1}_{N}\over h^{n}_{N+1}h^{n}_{N}}=1$. This can be
integrated once to
$h^{n+1}_{N}=h^{n}_{N}j(n-N)$ and then further integrated to
$h^{n}_{N}=J(n-N)\phi(N)$ such that $j(m)=J(m+1)/J(m)$.
Introducing the gauge $h$, equation (29) down-shifted once in $n$ becomes:
$$h^{n+1}_{N+1}h^{n-1}_{N}\tau^{n+1}_{N+1}{\tau}^{n-1}_{N}-2h^{n}_{N+1}
h^{n}_{N}\tau^{n}_{N+1}
\tau^{n}_{N}+\Bigl(1+{N+1-n\over
2\nu^2}\Bigr)h^{n-1}_{N+1}h^{n+1}_{N}\tau^{n-1}_{N+1}\tau^{n+1}_{N}=0
\eqno(32)$$
All the $\phi$'s factor out, and (32) coincides with (15a) provided a
single relation is
satisfied, which defines the gauge function $j$:
$$j(n-N-2)=j(n-N)\Bigl(1+{N+1-n\over 2\nu^2}\Bigr)\eqno(33)$$
Next we turn to (28), upshift it once in $n$, introduce $h$ and compare it
to (26). We remark that
thanks to (33) the ratio of the
$j$'s create exactly the right $n-N$-dependent terms in (26). We are left
with one condition that
defines $\phi$:
$${\phi(N+1)\phi(N-1)\over\phi(N)^2}={1\over 2\nu^2}\eqno(34)$$
Finally one can check that (30) is now identical to (27). Thus the two
bilinear formulations of
d-P$_{\rm II}$ are equivalent, provided one introduces the right gauge.

In this paper we have introduced a bilinear formalism for the description
of the
discrete Painlev\'e-II equation. This approach has turned out to be very
convenient for the
expression of the rational solutions that d-P$_{\rm II}$ possesses for
particular values of its
parameters. From the analysis presented above it is clear that the choice
of the appropriate gauge
for the description of the equations depends crucially on the type of the
solutions one wishes to
examine. The novel feature concerning the rational solutions of d-P$_{\rm
II}$ is that they can
be given in the form of Casorati determinants, a property that is not yet
fully explored even in
the continuous case. We expect that the bilinear formalism will be
extremely useful in the
study of the remaining discrete Painlev\'e equations and will help
establish a perfect parallel
between the discrete and the continuous case.

\noindent
\smallskip {\scap References}.
\smallskip
\item{[1]} P. Painlev\'e, Acta Math. 25 (1902) 1; B. Gambier, Acta Math. 33
(1910) 1.
\item{[2]} M.J. Ablowitz, and H. Segur, Phys. Rev. Lett. 38 (1977) 1103; H.
Flashka and A.C. Newell,
Com. Math. Phys. 76 (1980) 67; A.S. Fokas and X. Zhou, Comm. Math. Phys.
142 (1991) 313.
\item{[3]} V.A. Gromak and N.A. Lukashevich, {\sl The analytic solutions of
the Painlev\'e
equations}, (Universitetskoye Publishers, Minsk 1990), in Russian.
\item{[4]} A.S. Fokas and M.J. Ablowitz, J. Math. Phys. 23 (1982) 2033.
\item{[5]} K. Kajiwara and Y. Ohta, in preparation.
\item{[6]} A. Ramani, B. Grammaticos and J. Hietarinta, Phys. Rev. Lett. 67
(1991) 1829.
\item{[7]} M. Jimbo, T. Miwa and K. Ueno, Physica D 2 (1981) 306.
\item{[8]} B. Grammaticos and A. Ramani, {\sl Discrete Painlev\'e
equations: derivation
and properties}, in {\sl Applications of analytic and geometric methods to
nonlinear differential
equations}, P.A. Clarkson ed.,  NATO ASI C413 (1993) 299.
\item{[9]} A. Ramani and B. Grammaticos, J. Phys. A: Math. Gen. 25 (1992)
L633.
\item{[10]} B. Grammaticos, F.W. Nijhoff, V. Papageorgiou, A. Ramani and J.
Satsuma,
Phys. Lett. A185 (1994) 446.
\item{[11]} K.M. Tamizhmani, B. Grammaticos and A. Ramani, Lett. Math.
Phys. 29 (1993) 49.
\item{[12]} K. Kajiwara, Y. Ohta, J. Satsuma, B. Grammaticos and A. Ramani,
J. Phys. A 27 (1994)
915.
\item{[13]} J. Hietarinta and M.D. Kruskal, {\sl Hirota forms for the six
Painlev\'e equations
from singularity analysis}, in {\sl Painlev\'e transcendents: their
asymptotics  and physical
applications}, D. Levi and P. Winternitz eds, NATO ASI series B278, Plenum
1992, p.175.

\end